\documentclass{article}
\usepackage{frascatiphys}
\usepackage{graphicx}
\begin{document}
\title{
UHECR AND GRB NEUTRINOS:\\
AN INCOMPLETE REVOLUTION?
}
\author{
Daniele Fargion, Phys.Depart. Rome Univ.1 and INFN Rome1,\\
 Sapienza, Ple. A. Moro 2, Rome, 00185, Italy       \\
}
\maketitle
\baselineskip=11.6pt
\begin{abstract}
 At highest energy edges   Ultra High Energy Cosmic Ray, (UHECR) and PeVs neutrino (UHE$\nu$), should soon offer  new exciting astronomy. The fast and somehow  contradictory  growth of hundred of antagonist models shows the explosive vitality of those new astronomy frontiers. No conclusive understanding on the UHECR and UHE neutrino source are at hand.  The earliest expectation of GRBs (as a one shoot Fireball model) as the main (UHE$\nu$) sources has been rejected. The source of UHECR as the expected  GZK ones within our Super-Galactic Plane (within few tens Mpc) it has been quite disproved. However alternative models on GRB (as the long life precessing Jets) and the new updated records by AUGER, TA, ICECUBE are offering nevertheless partial understanding and early hint for point source correlations along our galaxy and toward Cen A, the nearest extragalactic AGN.
\end{abstract}
\baselineskip=14pt

\section{The Cosmic Ray Century }
\begin{figure}[htb]
    \begin{center}
        {\includegraphics[scale=0.59]{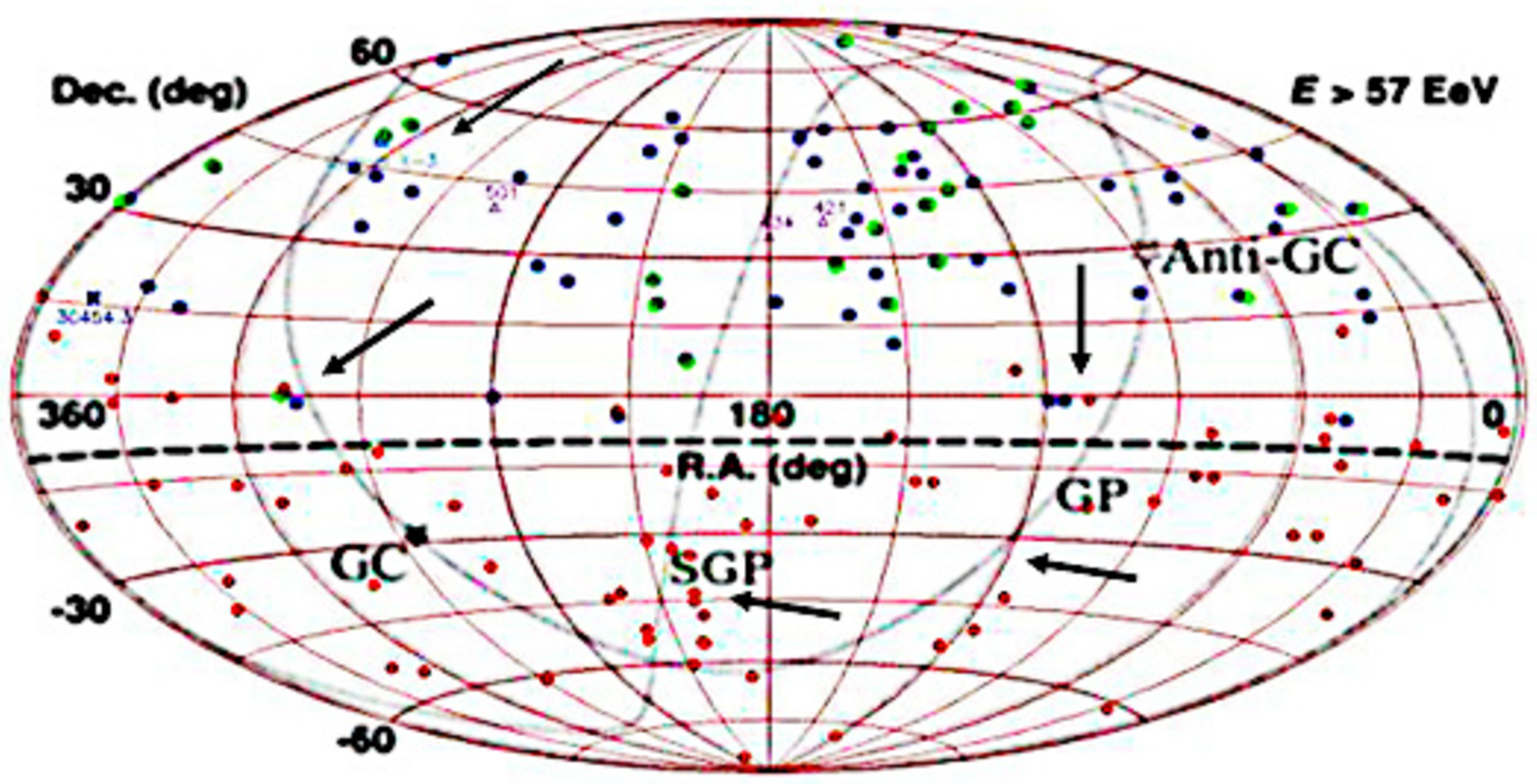}}\hspace{0.5cm}
  \caption{\it The UHECR map  in equatorial  coordinate following AUGER detector (South) and TA (in North)\cite{Olinto}. There are few clustering  tagged by five arrows: the most left and north one point toward a TA-Hires multiplet along Cen-3X,  one of the location of gamma anisotropy at TeV by Milagro-Argo detectors; the next arrow point toward the remarkable UHECR  made by highest doublet (TA-AUGER) UHECR found recently \cite{Troizky} where a new third UHECR has been just found this year by TA; the probability to occur such an event  is well below $10^{-4}$ ; a third arrow  on the center-right side point on to a clustering along a nearest AGN, Cen-a  not far from the largest energetic 2 PeV event number n.$35$ in ICECUBE over a doublet around the Cen A cluster,  where an additional twin overlapping multi-plet occurs at $20$ EeV energy  see \cite{AUGER11}; the fourth arrow point to the nearest and brightest Gamma Pulsar, Vela, related to an aligned AUGER triplet event; the fifth arrow point to a doublet by TA and a singlet by AUGER, an additional single by Hires  UHECR events nearby a well collimated  UHE neutrino muon found by ICECUBE  event n.$5$; all these  five region are candidate of UHECR and possibly UHE neutrino sources, mostly galactic.}
\label{exfig}
    \end{center}
\end{figure}

A century ago radioactivity was used to probe atomic nature. The same radioactivity was
discovered around us made by electrons (beta), alfa (Helium nuclei) and gamma (photons).
Radioactivity was apparently mainly made by terrestrial matter; indeed as one rise far from
the soil radioactivity start to decline.
However Victor Hess, a century ago, discovered that at highest altitude  while in balloons
 the radioactivity first decrease , but soon, (a few km above the sea level), it grows
 to tens or even hundred times more than sea level: a cosmic  radioactivity rules at highest  sky,
 consequently the cosmic rays (CR) nature was born. To day we do know that CR are mixed up but they are
  mostly charged particles as proton, Helium and other nuclei; CR are well representative
  of our solar system element composition; electron and positron are also present in CR.
  Photons and Neutrinos are shining too. Photons (gamma from X to TeVs energy) are well
  observed. Neutrinos not, mainly because CR atmospheric neutrino
  secondary pollution and because extreme neutrino weak interaction. Cosmic Ray energies ranges in an almost steady
 power law for nearly eleven order of magnitude from GeV to ZeV energy. Cosmic rays are mostly stop at twenty km. altitude
 (ten meter water equivalent) by our safe protective terrestrial atmosphere. Moreover
 Charged CR are smeared by terrestrial, solar and galactic  magnetic fields leading to confused
 homogeneous rain with no apparent source imprint. We are  blind within such a smooth CR rain. No source, no astronomy at sight.
 In some sense the existence of large scale galactic magnetic fields (smearing CR)  are testimony of the (mysterious) cosmic
 magnetic monopole absence (the celebrated Parker monopole bound).
 To be more accurate here on sea level we feel only a part of such smooth CR secondaries, traces made by scattering fragments
 of nucleons-nuclei that are raining in high altitude atmosphere, fragments known as muons or gamma and electron pairs, as well as
secondary neutrino, called because of it, atmospheric neutrinos. These are the neutrino noises hiding a more rare underline neutrino astronomy.
Most power-full CR (TeVs-PeVs-EeVs-ZeVs) are observed on  sea level by their catastrophic pair-production chain, leading to a tree-like air-shower whose top vertex is the primary CR event and whose late ramification are the
million or thousand of billion secondaries pairs leptons (and few hadrons). To observe such UHECR there are both water Cherenkov array detector (surface detectors in $km^2$ array) or Fluorescence Telescope array tracking air-shower lightening in the dark nights (in AUGER,HIRES,TA) experiments.
At TeVs-PeVs energies  CR air showering might blaze by Cherenkov flash, large optical telescope arrays (as Hess, Veritas, Magic)  or large scintillator and water arrays as ARGO and Milagro as well as ice (ICECUBE). These telescope experiments found in last decade a large number of TeVs point sources (partially coincident with $\gamma$ GeVs Fermi satellite signals); the array ones,ARGO and Milagro and ICECUBE CR, found a remarkable (tens degree size) anisotropy in the TeVs sky whose source is puzzling; we suggested  an UHECR radioactive beamed imprint (mostly galactic) and its decay in flight as a possible source \cite{Fargion-2011b}. We also try to find here UHECR and UHE neutrino correlation as discussed and summarized in Fig \ref{exfig}.
 Charged particles in CR at higher and higher energies maybe accelerated on star flare, Supernova explosion shells, jets either in micro (as GRBs, SGRs) and-or in macro sites (as AGN nuclei, Quasars), in brightest Radio Galaxies or even along more candidate places \cite{Merzaros}. We are all hunting for the CR sources nature since a century with no definitive success (yet). UHECR and UHE neutrino might point to them. Two main UHE neutrino models rise in last a few years: a Galactic and  an Extragalactic origination, each of the two defending the UHECR and UHE neutrino traces within different arguments.  Indeed in last decade  we all hoped to reveal soon their origination by their extreme  UHECR  (possibly un-deflected) component, hundred billions times more energetic than lower GeV-TeV CR. UHECR if nucleon while crossing in random walk in our $\mu$ Gauss magnetic fields, have to fly within narrow angle $\delta_{rm-p}\simeq 2.5^{o}$. No such a clustering multiplet, out of a remarkable a triplet to be discussed, has been found see Fig\ref{exfig}.  UHECR, following AUGER composition data might be nuclei, light ones (He-like); in that case the incoherent random angle bending  along the galactic plane and arms,  crossing along the whole Galactic disk of 20 kpc arriving  in different (alternating) spiral arm fields and within a characteristic coherent length of 2kpc for He nuclei becomes $\delta_{rm-He}\simeq 16^{o}$ \cite{Fargion-2011b}, well consistent with observed UHECR Cen A clustering, see Fig\ref{exfig}. On the contrary UHECR might be heavy (Fe,Ni..) ones, explaining the extreme spread of UHECR events along Vela that is the nearest and brightest Gamma pulsar in the sky,  see Fig\ref{exfig}.

 \section{The one shoot Fireball failure versus GRB precessing jet}
  The UHE $\nu$ signature was expected
   to be associated in time with several GRB (Gamma Ray Burst) sources or with BL Lac flaring sources ; other source candidate are the Star-Burst galaxy and the Radio Galaxies. The most popular Fire-ball model of GRB (one huge shoot event, within or without a fountain jet) has been disproved. It should be noted that if GRB are not one shoot event but a long  (decaying in hours-day scale time$ (\frac{t}{10^{4} s})^{-1}$ life)  jet \cite{Fargion99} than its precessing blazing time (for neutrinos) should not be  longer correlated with the short gamma-X blazing time (a constraint assumed of second-minute in Fireball times). Indeed in precessing GRBs Jet model it has been assumed since long ago a wider time scale to embrace also (otherwise mysterious) GRBs (as well as SGRs) precursors and statistical Jet solid angle \cite{Fargion99} \cite{Fargion05}. These observed precursor may be both gamma and  neutrino events; such precursor neutrinos are indeed observed in a few GRB events: a $109$ TeV neutrino, within $0.2^{o}$ of GRB091230A, with a localization uncertainty of $0.2^{o}$, and detected time some $14$ hours before the gamma GRB091230A trigger; a $1.3$ TeV neutrino $1.9^{o}$ off GRB090417B, with localization uncertainty of $1.6^{o}$, and detection time $2249$ seconds before the trigger of GRB090417B; a $3.3$ TeV neutrino $6.1^{o}$ off GRB090219, with a localization uncertainty of $6.1^{o}$, and detection time $3594$ seconds before the GRB090219 trigger.
    These three observed GRB-neutrino precursor neutrino event (unexplained  in fireball one-shoot model) may be, at a few percent of all the UHE neutrino events in ICECUBE. Additional GRBs might be at higher redshift and they may be part of the ICECUBE UHE neutrino sources if one enlarge the GRB-neutrino time windows; however also AGN jet and local galactic UHE $\nu$ may play a role. As we shall show a  few correlated UHE neutrino and UHECR may be present inside galactic plane as well as few smeared clustering in tens TeV $\gamma$ CR anisotropy and spread multiplet in tens EeV UHECR around Cen-A may trace the UHECR astronomy, see details in  Fig.\ref{exfig}. It should be noted that if UHE neutrino will cluster in a spread tail group of events one might advocate  either  UHECR scattering along galactic gas (as for the Fermi bubble traced by its observed $\gamma$ fountain), or one may suggest the radioactive (light and heavy) UHECR decay in flight, offering a possible correlation to UHECR $10^{18}-10^{20}$ eV clustering and the large scale Milagro-Argo-ICECUBE TeVs-PeVs $\gamma$ anisotropy \cite{Fargion-2009}, \cite{Fargion-2011b}.
 \section{The UHE neutrino flavor metamorphosis}
    Let us remind that neutrino are neutral and always un-deflected; (to be more precise in cosmology and for neutrino with mass  in expanding universe once they became non relativistic  neutrino might be bent by gravity too; this  may play a minor role in largest scale dark matter density growth and large scale formation). Therefore the UHE neutrino may offer a new Astronomy.
     However, as we mentioned, at low (less than tens TeV) energies, neutrinos  are polluted by abundant (CRs fragments),
     a smeared atmospheric neutrino noise that are  hiding any underlying astrophysical neutrino point-source.
     Let us remind that atmospheric GeV neutrinos while being mostly born at a  ratio
      ($\nu_{e}:\nu_{\mu}:\nu_{\tau}$= $1:2:0$)  by proton-proton scattering  interactions chain in atmosphere,
       once at hundred GeV-TeVs,  they become ruled, at sea level, by muons neutrinos
($\nu_{e}:\nu_{\mu}:\nu_{\tau}\simeq$ $0.1:1:0$); this occurs because the relativistic pion (and Kaon) still decay,
 feeding muons and anti-muons neutrinos while their muon decay, the main road to electron flavor birth, is inhibited   by a longer muon lifetime.
Therefore electron (and rarest charmed born tau flavor)  presence  at hundred GeV-TeVs are rare (less than $10\%$);
 most  signals (at hundred GeV-TeVs) are muon neutrino tracks
as observed by inner ICECUBE experiment, the Deep Core, in the last years.
    Here we don't  discuss the atmospheric muon neutrino conversion and partial suppression by flavor mixing
     that is tuned at GeV energies and led to the Pontecorvo–Maki–Nakagawa–Sakata  in last two decades.
     To consider the flavor mixing and possible experiment along the Earth see \cite{Fargion-2012}
     However at higher energies (above tens TeV-PeV) the  extraterrestrial signals might (and indeed  do)  overcome  the softer atmospheric neutrino noise (mainly because the astrophysical  hardest spectra). Indeed those extraterrestrial signals, while being expected to be born, in general, by $p+p$ or $p + \gamma$ in flux ratio,($\nu_{e}:\nu_{\mu}:\nu_{\tau}\simeq $ $1:2:0$),
      because of the  mixing and because of the large galactic distances,
      may oscillate and converted  into  electron and  tau neutrino flavor component.
      The outcome  in a  first approximation lead to a final equipartition flavor flux:
       ($\nu_{e}:\nu_{\mu}:\nu_{\tau}$= $1:1:1$), because of the complete flavor de-coherence mixing in flight.
  These ruling shower signals have been observed  by last $3$ years ICECUBE
 event data. The dominant  $\nu_{e}$ and $\nu_{\tau}$  interaction are mainly electromagnetic
 leading to Cherenkov spherical shower in ice. These majority of spherical shower events (28)
 are four times more abundant than muon tracks (7). This may sound contradictory
 keeping in mind that a third of these 35 events should be  atmospheric muon or neutrino noise\cite{ICECUBE14} (in principle all of them noise signals should mainly show up as muons, too many respect $7$ observed ones); however because a very different detector flavor acceptance it maybe still consistent with ($\nu_{e}:\nu_{\mu}:\nu_{\tau}$= $1:1:1$) \cite{Vissani-2013},\cite{DFPP14}, once considering also the Neutral Current contribute.
 Therefore the best hope and probe  of a new UHE $\nu$ astronomy is the recent neutrino sudden   flavor change at tens TeV-PeV energy  \cite{DFPP14}. But all showering cascades are smeared in their arrival direction ($\pm 15^{o}$) and they are making inconclusive any map correlation. Therefore a more directional astronomy is needed, as the one made by muon tracks.  Additional such  signals are able to open an UHE neutrino astronomy : they are  in-written into few tens TeV energy ( crossing in ICECUBE) muons at horizons \cite{Fargion-2014}; the first estimate of nearly 40 of such events offer hope for soon novel astronomy.  Unfortunately the recent published (TeV-PeVs) (not only those above  tens-TeV-PeV)  muon crossing inside the ICECUBE were painting a puzzling random  map, not favoring any known or expected $\gamma$  X  source, \cite{ICECUBE14}.  A more restrictive filtering of those events (muon crossing above  few Tens TeV)  may be more telling for extraterrestrial nature, but such a selection has not been done yet \cite{Fargion-2014}. We must remind that an anti-neutrino electron $\bar{\nu_{e}} + e$ peak resonance may tag and reveal a different neutrino sky volume. Indeed a  Glashow resonance peak may rise by $\bar{\nu_{e}} + e \rightarrow W^{-}$ at $E_{\bar{\nu_{e}}} = 6.3 $ PeV \cite{glashow}, but it  has  not been observed (yet), while observing 2 PeV cascade shower. This is suggesting either a sudden softening in the PeV neutrino spectra  \cite{DFPP14} or a smooth interchange (around ten TeV energy) between a soft power law in atmospheric neutrino with a harder  extraterrestrial neutrino power law within a fine-tuned parameters (apt to avoid the expected Glashow signal as well as its ideal $\tau$ double bang signature)\cite{doub_bang}. The powerful  $\nu_{\tau}$  discover via its first bang inside a rock (mountain,
  or Earth) and  its consequent $\tau$ escape outside in air, decaying in an amplified $\tau$ airshower,   is a very promising adjoint neutrino astronomy at highest energy range PeVs-EeV, first foreseen  more than 15 years ago \cite{Fargion02} and to day widely searched in different large experimental array today \cite{Bertou2002}, \cite{AUGER13}, \cite{Aita2011},(the so called Earth skimming neutrinos, \cite{Feng02}).

 \section{The UHECR fly  undeflected: an UHECR Astronomy?}
 We expected that UHECR might flight straight  because of their energetic rigidity, at best for UHECR proton. They survive the Lorentz bending and smoothing occuring for lower (up to EeV)  CR. Such UHECR, well above EeV, are extremely rare, but their interaction at high altitude in atmosphere makes their explosive pair-production tree air-shower  along their  fall, proliferous and amplified into  extended  wide area (tens km square size).  Their detection may be tested by wide spread km distance array, each detector even of minor volume (few square meter swimming-pool for Cherenkov detection) on the ground. Such experiments  like Flys' Eye, AGASA, Hires, Auger, TA, were located  in last two decades over hundreds or several thousands km square area.  These UHECR (if nucleon or even nuclei) might exhibit a cut off (within nearly $2\%$ of cosmic radius)  because of the "Cosmic Back Ground Radiation" opacity, mostly by $p +\gamma \rightarrow \Delta \rightarrow \pi + nucleon$ interaction, the so-called photon-pion GZK cut off \cite{za66}. Their consequent GZK neutrinos (at EeV energies) are not observed yet and cannot feed the mentioned ICECUBE events. A more severe distance cut occur to UHECR nuclei propagation because of their fragility by photon-nuclei dissociation; moreover light and heavy nuclei are partially or totally bent while crossing the galaxy. Therefore  their maps might trace only nearby local Universe well within $1-2\%$ percent  of cosmic size.
 To escape the near Universe size (GZK size bounded, in case of UHECR clustering at far extragalactic edges) one may consider the UHE neutrinos at ZeV hitting the relic cosmic ones via Z-resonance (in analogy to Glashow W resonance) \cite{Fargion97}: the Z UHE decay into nucleons or antinucleons might be the final trace explaining UHECR correlation well above GZK bounded universe; this proposal found much interest and it will be actual if UHECR are correlated with guaranteed AGN above GZK distances.
  UHECR neutron are also expected but well confined within one Mpc ($E_{n}\simeq 10^{20}$ eV) size while UHE photons ($E_{\gamma}\simeq 10^{18}-10^{20}$ eV) are bounded within a few tens Mpc. No such UHECR neutron or photon source or clustering has been found (yet).  UHE neutrino might test most of the far and secret Universe edges, but they may observe also nearby galactic sources. The simplest solution of UHECR (nearby Local Super-galactic plane) and of UHE neutrino (expected to be traces of GRBs) have been in a very recent years  fallen away. New galactic and extragalactic candidate source have been considered, somehow with much dispute and  disagreement in the scientific arena. Therefore UHECR either nucleon or nuclei must arise in a small (tens Mpc) or even narrow (few Mpc) Universe, possibly in sharp astronomy (for proton) or in a smeared clustering map (for light nuclei or nearest galactic heavy nuclei). The last AUGER maps showed only marginal smeared clustering and a rarest remarkable triplet \cite{Troizky}. see Fig.\ref{exfig}.
  \section{Conclusions}
The difficult puzzle of UHECR astronomy and the UHE neutrino maps may soon be matched by cooperative test and overlapping. There are often unexplainable delay in UHECR (AUGER) data release.  Nevertheless the sources as nearest AGN Cen A, rise as a remarkable smeared clustering  of UHECR events in AUGER ($E_{UHECR} > 6 \cdot 10^{19}$ eV as well in rare overlapping tens EeV long chain events foreseen \cite{Fargion-2009}, and observed,\cite{AUGER11} \cite{Fargion-2011b}, may be well understood if they are mostly made by He nuclei and its fragments.  The nearest brightest $\gamma$ pulsar  Vela is also suspected to correlate a train of UHECR events (if heavy Fe,Ni,Co nuclei), and a doublet of ICECUBE neutrinos ( see Fig.\ref{exfig}, event n. 3 (a muon) and 6 (a shower) in ICECUBE \cite{ICECUBE14a})as well as a remarkable TeVs ICECUBE CR anisotropy \cite{Fargion-2011b}, Cen-X3 and Cygnus region is also rising in ARGO TeVs anisotropy \cite{Fargion-2011b} and in a very rich ($7$) recent UHECR multiplet clustering containing also new TA and old Hires events. The most surprising narrow triplet is the newest rarest  highest energy doublet \cite{Troizky}, by  an additional third event (by last  TA UHECR data); other  triplet and quadruplet point to unknown sources not far from galactic plane; they are possibly showing  a cooperative galactic and extragalactic source role; we offered here  first attempts in this difficult map  understanding (see Fig.\ref{exfig}). We believe that with  care and with needed time we are going to disentangle ( within the fog of such noisy high energy sky) the first sources shining from our near and far Universe; we believe that most are related to precessing jet beaming, galactic and extragalactic in tuned and equiparable ratio (see also \cite{Padovani-2014}), and UHECR as well as UHE neutrino are not found along contemporaneous explosive or flaring event, because of a different timing of the jet blazing beam. UHE neutrinos are by most distant GRB and AGN blazing whose  timing maybe often delayed or precursor respect gamma flaring or burst. On the contrary UHECR are mostly galactic or in nearest Universe. Few sources (UHECR-UHE $\nu$)  overlap within nearest galaxy and AGN sky.
 \section{In memory}This article is devoted to Daniele Habib greatest linguist and translator, 
 who disappeared in these days half a century ago.

\end{document}